\documentclass[a4paper,11pt]{article}
\pdfoutput=1 

\usepackage{jheppub} 
\usepackage{amssymb,amstext,amsthm,amsfonts,mathrsfs,amsbsy,amsmath,array,multirow}
\usepackage[T1]{fontenc} 

\rightline{\small IPhT-T14/066\hspace*{-2cm}}

\title{\boldmath Regular 3-charge 4D black holes and their \\
microscopic description}

\author{Iosif Bena and}
\author{C. S. Shahbazi}

\affiliation{Institut de Physique Th\'eorique, CEA Saclay.}

\emailAdd{iosif.bena@cea.fr}
\emailAdd{carlos.shabazi@cea.fr}

\abstract{\vspace{0.1cm}\\ The perturbative $\alpha^{\prime}$ corrections to Type-IIA String Theory compactified on a Calabi-Yau three-fold allow the construction of regular three-charge supersymmetric black holes in four dimensions, whose entropy scales with the charges as $S \sim \left( p^1 p^2 p^3\right)^{\frac{2}{3}} $ . We construct an M-theory uplift of these ``quantum'' black holes and show that they can be interpreted as arising from three stacks of M2 branes on a conical singularity. This in turns allow us relate them via a series of dualities to a system of D3 branes carrying momentum and thus to give a microscopic interpretation of their entropy.}

\begin{document}
\maketitle
\flushbottom


\section{Introduction}


String theory has proven to be extremely successful in reproducing the entropy of supersymmetric black holes, such as the three-charge black hole in five dimensions \cite{Strominger:1996sh} and the four-charge black hole in four dimensions \cite{Johnson:1996ga,Maldacena:1996gb,Maldacena:1997de}. The entropies of these black holes scale like the square root of the product of their charges (or some some duality-invariant form thereof \cite{Kallosh:1996uy}), and in the microscopic counting this square root comes from using Cardy's formula to count the states of a certain 1+1 dimensional system of strings and branes. 

However, in certain four-dimensional compactifications of string theory one can construct three-charge black holes whose entropy scales with these charges like  $S \sim \left( p^1 p^2 p^3\right)^{\frac{2}{3}} $ \cite{Bueno:2012jc}. The curvature at the horizon of these black holes is small, precisely as one would expect from the fact that their entropy grows like the square of the charge. These black holes cannot be constructed in ``normal'' four-dimensional supergravity, where the horizon curvature of three-charge black hole is Planckian, but exist if one adds to the supergravity Lagrangian certain terms coming from perturbative String-Theory  $\alpha^{\prime}$ corrections  at tree level in $g_{S}$.  Since these black holes do not have a regular limit when these correction terms are removed \footnote{Unlike other black hole solutions constructed in these $\alpha^{\prime}$-corrected theories\cite{Behrndt:1997gs,Behrndt:1997ei,Gaida:1997id,Gaida:1998pz}} they are called ``quantum black holes.'' 

The purpose of this letter is to try to understand the microscopic entropy of the type IIA quantum black holes constructed in \cite{Bueno:2012jc}. The first step in this direction is to uplift these black hole solutions to eleven dimensions and to propose an M-theory interpretation in terms of three mutually-orthogonal intersecting stacks of M2-branes at the tip of a four-dimensional Gibbons-Hawking-like base. Since the $\alpha^{\prime}$ corrections to Type-IIA Calabi-Yau compactifications come from higher-derivative terms in String Theory \cite{Antoniadis:1997eg,Antoniadis:2003sw}, the uplifted black hole will be a solution of the equations of motion of eleven-dimensional supergravity modified by the addition of certain higher-derivative terms. 

The second step is to argue that if the Calabi-Yau manifold can be written as a (possibly singular) elliptic fibration, the branes that make the eleven-dimensional solution can be dualized to a configuration of two intersecting mutually-supersymmetric D3 branes carrying momentum along their common direction. By counting the possible way of carrying this momentum and by remembering that these D3 branes sit on top of a conical singularity that effectively enhances the central charge of the 1+1 dimensional theory on their worldvolume, we are able to reproduce the peculiar charge dependence, $S \sim \left( p^1 p^2 p^3\right)^{\frac{2}{3}} $, of the entropy of the quantum black holes. 

In section \ref{sec:IIAquantumblackholes} we introduce the effective theory corresponding to Type-IIA String Theory compactified on a Calabi-Yau manifold as well as the corresponding Type-IIA quantum black hole solutions. In section \ref{sec:Mth} we argue that the M-theory uplift of quantum black holes can be interpreted as arising from three stake of intersecting M2 branes and in section \ref{sec:microscopic} we use this to propose a microscopic description of their entropy. Section \ref{sec:conclusions} contains conclusions and future directions.  In the appendices (\ref{sec:N2SugraHFGK}) and (\ref{sec:quantumblackholes}) we review the construction of Type-IIA quantum black holes as well as the H-FGK formalism \cite{Mohaupt:2010fk,Mohaupt:2011aa,Meessen:2011bd,Galli:2011fq,Galli:2012ji,Meessen:2012su,
Galli:2012pt,Bueno:2013pja,Bueno:2013psa} and the structure of $\mathcal{N}=2$ four-dimensional ungauged supergravity coupled to vector multiplets that were used in their construction.


\section{Type-IIA Quantum Black Holes}
\label{sec:IIAquantumblackholes}


Quantum black holes \cite{Bueno:2012jc} are solutions of the effective theory corresponding to Type-IIA String Theory compactified on a Calabi-Yau manifold in the presence of perturbative corrections to the Special K\"ahler geometry of the vector multiplet sector. Despite having only three charges, these four-dimensional supersymmetric black holes have a macroscopically large horizon area. It is not hard to see either from this or from their explicit construction that these black holes do not have a macroscopic horizon in the ``classical'' limit, when the perturbative corrections are turned off, which justifies calling them ``quantum black holes''.


\subsection{Type-IIA String Theory on a Calabi-Yau manifold}
\label{sec:iiacy}


\hspace{0.4cm} Type-IIA String Theory compactified to four-dimensions on a Calabi-Yau three-fold, with Hodge numbers $(h^{1,1},h^{2,1})$, is described, up to two derivatives, by a $\mathcal{N}=2$ four-dimensional supergravity coupled to vector- and hyper-multiplets. As explained in appendix (\ref{sec:N2SugraHFGK}), we are going to truncate the hyperscalars to a constant value, and therefore we will only be concerned about the vector-multiplet sector of the theory. The corresponding prepotential can be written as an infinite series around $\Im{\rm m}z^i\rightarrow \infty$\footnote{Actually, the prepotential obtained in a Type-IIA Calabi-Yau compactification is \emph{symplectically equivalent} to the prepotential (\ref{eq:IIaprepotential}).}, and it is given by \cite{Candelas:1990pi,Candelas:1990rm,Candelas:1990qd,
Mohaupt:2000mj}

\begin{equation}
\label{eq:IIaprepotential}
\mathcal{F} =  -\frac{1}{3!}\kappa^{0}_{ijk} z^i z^j z^k +\frac{ic}{2}+\frac{i}{(2\pi)^3}\sum_{\{d_{i}\}} n_{\{d_{i}\}} Li_{3}\left(e^{2\pi i d_{i} z^{i}}\right) \ \, ,
\end{equation}

\noindent
where $z^i,~~ i =1,...,n_v=h^{1,1}$, are the scalars in the vector multiplets, $\kappa^{0}_{ijk}$ are the classical intersection numbers, $d_{i}\in\mathbb{Z}^{+}$ is a $h^{1,1}$-dimensional summation index and $Li_{3}(x)$ is the third polylogarithmic function. The constant $c$ is proportional to the Euler characteristic of the Calabi-Yau three-fold, $\chi$ multiplied by the Riemann zeta function: $c\equiv\frac{\chi\zeta(3)}{ (2\pi)^3}$. 

The first two terms in the prepotential correspond to tree level and fourth-loop perturbative (which is the only non-vanishing one) contributions in the $\alpha^{\prime}$-expansion, respectively \cite{Mohaupt:2000mj}

\begin{equation}
\label{eq:IIapert}
\mathcal{F}_{\text{P}} =  -\frac{1}{3!}\kappa^{0}_{ijk} z^i z^j z^k +\frac{ic}{2}\ \, ,
\end{equation}

\noindent
while the third term comes from non-perturbative corrections produced by world-sheet instantons. 
In this paper we will focus on large-volume compactifications where these corrections can be safely ignored, and hence focus on the quantum black holes obtained from the prepotential (\ref{eq:IIapert}). The most general quantum black hole solutions, that are governed by the prepotential  (\ref{eq:IIaprepotential})  have been constructed in \cite{Bueno:2013psa}.

Our starting point is the prepotential (\ref{eq:IIapert}), which in homogeneous coordinates $\mathcal{X}^{\Lambda}, ~~\Lambda=(0,i)$, can be written as

\begin{equation}
\label{eq:prepotential}
F(\mathcal{X})=  - \frac{1}{3!}\kappa^{0}_{ijk}  \frac{\mathcal{X}^{i}\mathcal{X}^{j}\mathcal{X}^{k}}{\mathcal{X}^0} +\frac{ic}{2} \left(\mathcal{X}^0\right)^2 \ \, .
\end{equation}

\noindent
The scalars $z^i$ are given by

\begin{equation}
\label{eq:scalarsgeneralII}
z^i=\frac{\mathcal{X}^{i}}{\mathcal{X}^0} \ \, .
\end{equation}

\noindent
Adding the constant term $c$ to the prepotential modifies the geometry of the scalar manifold, which  is no longer homogeneous, and therefore it is said that the geometry has been \emph{corrected} by quantum effects. The scalar geometry defined by (\ref{eq:prepotential}) is hence the so-called \emph{quantum corrected} $d$-SK geometry \cite{deWit:1991nm,deWit:1992wf}. The attractor points of (\ref{eq:prepotential}) have been extensively studied in \cite{Bellucci:2009pg}. 

The Type-IIA quantum black holes belong to a particular class of purely magnetic black hole solutions of the theory defined by (\ref{eq:prepotential}). We review them in appendix (\ref{sec:IIAquantumblackholes}) and refer the reader to \cite{Bueno:2012jc,Bueno:2013psa} for more details. 


\subsection{The quantum-corrected STUmodel}
\label{sec:STUblackholes}


Type-IIA quantum black holes exist in any Type-IIA Calabi-Yau compactification with $h^{1,1}>h^{2,1}$. To make contact to a description of these black holes in terms of intersecting branes we have to choose a particular model and the best candidate is the popular STU model, whose black hole solutions and attractors have been extensively studied. Since the hypermultiplets are truncated, the specific value of $h^{2,1}$ is irrelevant as long as it is smaller than $h^{1,1}$. The values of $h^{1,1}$ and $\kappa^{0}_{ijk}$ are

\begin{equation}
\label{eq:choice}
h^{1,1}=1\, , \quad\kappa^{0}_{123}=1\, 
\end{equation}

\noindent
and therefore the prepotential (\ref{eq:prepotential}) becomes

\begin{equation}
F\left(\mathcal{X}\right) = \frac{\mathcal{X}^{1}\mathcal{X}^{2}\mathcal{X}^{3}}{\mathcal{X}^{0}} + \frac{ic}{2} \left(\mathcal{X}^{0}\right)^2\, .
\end{equation}

\noindent

The ``normal'' four- and five-dimensional black holes and attractors of this quantum-corrected STU model been previously considered in \cite{Gaida:1997id,Behrndt:1997ei,Behrndt:1997gs,Gaida:1998pz,Bellucci:2009pg,Galli:2012pt}, but here we focus on the black holes that do not have a classical ($c\rightarrow 0$) limit. The solution corresponding to these black holes (discussed in detail in Appendix  \ref{sec:quantumblackholes}) has

\begin{equation}
\label{eq:hessianSTU}
e^{-2U} = 3 c^{\frac{1}{3}}\left|H^{1} H^{2} H^{3}\right|^{2/3}\, ,
\end{equation}

\noindent
\begin{equation}
\label{eq:scalarsSTU}
z^i= i c^{\frac{1}{3}} \frac{H^i}{\left(H^{1} H^{2} H^{3}\right)^{1/3}} \, .
\end{equation}

\noindent
The space-time metric is therefore 

\begin{equation}
\label{eq:generalbhmetricsusySTU}
\begin{array}{rcl}
ds^2_{4} = 
-3^{-1}c^{-\frac{1}{3}}\left|H^{1} H^{2} H^{3}\right|^{-2/3} dt^2 + 3c^{\frac{1}{3}}\left|H^{1} H^{2} H^{3}\right|^{2/3} d\vec{y}^2 \, ,
\end{array}
\end{equation}

\noindent
where $d\vec{y}^2 = \left( dy^{1}\right)^2+\left( dy^{2}\right)^2+\left( dy^{3}\right)^2$ is the Euclidean metric on $\mathbb{R}^{3}$. Since in the H-FGK formalism the $H$-variables correspond to the imaginary part of the covariantly holomorphic symplectic section appropriately weighted to be K\"ahler neutral, supersymmetry require them to be harmonic functions on the transverse space $\mathbb{R}^{3}$. A single-center black hole has then 

\begin{equation}
H^{i} = a^{i}+\frac{p^{i}}{\sqrt{2}}\frac{1}{r}\, , \quad i =1,2,3\, ,
\end{equation}

\noindent
where $r^2 = \left( y^{1}\right)^2+\left( y^{2}\right)^2+\left( y^{3}\right)^2$, the $p^i$ are the three charges of the black hole (\ref{eq:chargesFG}) and the $a^{i}$ are arbitrary constants that can be written in terms of the asymptotic value of the scalars at spatial infinity $z^{i}_{\infty}$ as

\begin{equation}
a^{i} = -s_{p^{i}}\frac{\mathrm{Im}\, z^{i}_{\infty}}{\sqrt{3c}}\, ,
\end{equation}

\noindent
where $s_{p^{i}}$ is the sign of the charge $p^{i}$. The entropy of this black hole is 

\begin{equation}
\label{eq:entropySTU}
S = \frac{3c^{\frac{1}{3}}}{2} \pi\left| p^1 p^2 p^3\right|^{2/3}\, 
\end{equation}

\noindent
and its mass is the sum of the three charges:

\begin{equation}
M =\sqrt{\frac{3c}{2}}\left(a_2 a_3 p^1 + a_1 a_3 p^2 + a_1 a_2 p^3\right)\, .
\end{equation}

\noindent
It is easy to see that each term that contributes to the mass is positive definite since $\mathrm{Sign}\, a^{i} = \mathrm{Sign}\, p^{i}$ and $a^1 a^2 a^3 > 0$. In the next section we will try to describe this three-charge four-dimensional black hole in terms of intersecting branes in M-theory.
 

\section{The M-theory configuration}
\label{sec:Mth}


In order to see whether Type-IIA quantum black holes have an interpretation via intersecting branes it is desirable to have the precise ten-dimensional configuration corresponding to the four-dimensional solution. For tree-level Type-IIA Calabi-Yau compactifications the map between the ten-dimensional and the four-dimensional fields is known \cite{Ferrara:1988ff,Cecotti:1988qn,Ferrara:1989ik,Bodner:1990zm}, but since we are considering four-dimensional solutions to the prepotential that includes perturbative corrections in $\alpha^{\prime}$ at tree level in $g_{S}$, and since these correction come from  an $R^{4}$-like term in ten dimensions, no explicit map is known. However, as we will show below, we will still be able to control the behavior of the dilaton and the Calabi-Yau volume, which will allow us to obtain the higher-dimensional configuration corresponding to Type-IIA quantum black holes.


\subsection{The String Theory dilaton}
\label{sec:dilaton}


As explained in appendix (\ref{sec:N2SugraHFGK}), for black hole solutions of ungauged four-dimensional supergravity, the hyperscalars are truncated to a constant value. In principle, the dilaton belongs to the universal hypermultiplet, and therefore one may naively conclude that it should be constant for every black hole solution. However, in the process of obtaining the effective $\mathcal{N}=2$ four-dimensional supergravity in its standard form (\ref{eq:action}), several rescalings and redefinitons are performed on the original ten-dimensional fields. In addition, since we are considering all the perturbative corrections to the Special K\"ahler sector, the corresponding $\mathcal{N}=2$ ungauged supergravity action is not the effective compactification theory of Type-IIA String Theory at tree level, so we should expect more intrincate redefinitions. Our purpose is to show now that Type-IIA quantum black holes have a constant ten-dimensional dilaton and a constant Calabi-Yau manifold volume, and we will do this in two steps:

\paragraph{Tree level:} At tree level, the Type-IIA dilaton $\phi$ is related to the four-dimensional dilaton $q$ as follows \cite{Gurrieri:2003st,Grimm:2005fa}

\begin{equation}
e^{-q} = e^{-\phi} \sqrt{\mathrm{Vol}_{6}}\, ,
\end{equation}

\noindent
where

\begin{equation}
\mathrm{Vol}_{6} = \frac{1}{6}\int J\wedge J \wedge J\, ,
\end{equation}

\noindent
is the volume of the Calabi-Yau manifold, $J$ being the corresponding K\"ahler form. The K\"ahler potential of the $\mathcal{N}=2$ Special K\"ahler manifold is related to the volume of the compactification Calabi-Yau manifold by \cite{Gurrieri:2003st,Grimm:2005fa}

\begin{equation}
e^{-\mathcal{K}} = \frac{e^{\frac{3\phi}{2}}}{6}\int J\wedge J \wedge J = e^{\frac{3\phi}{2}} \mathrm{Vol}_{6}\, .
\end{equation}

\noindent
Since for Type-IIA quantum black holes both $e^{-\mathcal{K}}$ and $q$ are constants, this implies that both

\begin{equation}
 e^{-\phi} \sqrt{\mathrm{Vol}_{6}} = {\rm const.} ~\quad~ {\rm and}~\quad~ e^{\frac{3\phi}{2}} \mathrm{Vol}_{6}  = {\rm const.}\, 
\end{equation}

\noindent
and hence at tree level both the dilaton $\phi$ and the Calabi-Yau volume are constant.

\paragraph{Perturbative corrections:} It is also easy to see that this tree level result is not changed when including perturbative corrections. Indeed, \cite{Antoniadis:1997eg,Antoniadis:2003sw}  have shown that the loop corrections in ten dimensions that give rise to the perturbative corrections of the prepotential from the four-dimensional point of view, only mix the dilaton and the volume among themselves. Therefore, since they were constant at the tree level, they continue to be constant after the loop corrections have been taken into account. We thus conclude that for Type-IIA quantum black holes the dilaton and the volume of the Calabi-Yau manifold are constant.


\subsection{M2-branes}


In order to argue that the M-theory uplift of quantum black holes can be interpreted as coming from a superposition of M2 branes on a conical singularity it is useful to recall the usual supersymmetric solution corresponding to three stacks of M2 branes on a six-torus \cite{Tseytlin:1996bh,Gauntlett:1996pb}

\begin{eqnarray}
\label{eq:M2M2M2electric}
 ds^2 = \left(H_{1}H_{2}H_{3}\right)^{\frac{1}{3}} \left[\, - \left(H_{1}H_{2}H_{3}\right)^{-1}dt^2 + H^{-1}_{1}\left(dx^{2}_{1}+dx^{2}_{2}\right)+H^{-1}_{3}\left(dx^{2}_{3}+dx^{2}_{4}\right)\right.\nonumber\\ \left.  +H^{-1}_{2}\left(dx^{2}_{5}+dx^{2}_{6}\right)+g_{\underline{m}\underline{n}} dy^{\underline{m}}dy^{\underline{n}}\,\right]\, ,
\end{eqnarray}

\noindent
where $\underline{m},\underline{n} = 1, \dots, 4$ and the M2-branes are respectively located along the $\left\{x^{1},x^{2}\right\}$, $\left\{x^{3},x^{4}\right\}$ and $\left\{x^{5},x^{6}\right\}$ directions. Supersymmetry requires the transverse metric, $g_{\underline{m}\underline{n}} $, to be Hyper-K\"ahler and the M2 warp factors to be harmonic in this metric: $\Delta_{y} H_{i}(y) = 0\, ,\,\, i = 1,2,3$.
To compare with the uplifted quantum black holes we will focus on solutions where the Hyper-K\"ahler space is a Gibbons-Hawking (Taub-NUT) space:
 
\begin{equation}
\label{eq:GH}
ds^2_{\mathrm{GH}} = V(y)^{-1}\left(d\Psi + A_{i}(y)dy^{i}\right)^2 + V(y)\delta_{ij} dy^{i} dy^{j}\, , \qquad i = 1,2,3\, ,
\end{equation}

\noindent
where 

\begin{equation}
\label{eq:GHconditions}
\ast dA(y) = dV(y)\,\, \Rightarrow\,\, \Delta_{y}\, V(y) = 0\, .
\end{equation}

\noindent
The eleven-dimensional metric is therefore given by

\begin{eqnarray}
\label{eq:M2M2M2electricII}
ds^2 = \left(H_{1}H_{2}H_{3}\right)^{\frac{1}{3}} \left[\,- \left(H_{1}H_{2}H_{3}\right)^{-1}dt^2 +H^{-1}_{1}\left(dx^{2}_{1}+dx^{2}_{2}\right)+H^{-1}_{3}\left(dx^{2}_{3}+dx^{2}_{4}\right) \right. \nonumber\\ \left. +H^{-1}_{2}\left(dx^{2}_{5}+dx^{2}_{6}\right)+ds^2_{\mathrm{GH}}\,\right]\, , 
\end{eqnarray}

\noindent
where now the $H_{i}$ are harmonic in $\mathbb{R}^3$.


\subsection{M2-branes beyond classical supergravity}
\label{sec:quantumM2}


In order to interpret Type-IIA quantum black-holes as composed of three stacks of orthogonally intersecting M2-branes, one should uplift their solution to M-theory. Since the four-dimensional theory where these black holes are constructed includes all the perturbative corrections in $\alpha^{\prime}$, at tree level in $g_{S}$, the uplifted black holes should be solutions of the standard eleven-dimensional supergravity to which one has added the higher-derivative terms that give rise to the 4D perturbative $\alpha^{\prime}$-corrections. 

There are two features of the quantum black hole metric (\ref{eq:generalbhmetricsusySTU})  that will guide us to obtain this solution. The first is that the volume of the six-dimensional torus/Calabi-Yau manifold is constant to all levels in the corrections and hence, rescaling the time to $t\to  \sqrt{3 c^{\frac{1}{3}}} t\,$, the eleven-dimensional solution can be put into an M2-brane form:

\begin{eqnarray}
\label{eq:M2M2M2electricAnsatz}
ds^2 = \left(H_{1}H_{2}H_{3}\right)^{\frac{1}{3}} \left[\, -\left(3c^{\frac{1}{3}} H_{1}H_{2}H_{3}\right)^{-1} dt^2 + H^{-1}_{1}\left(dx^{2}_{1}+dx^{2}_{2}\right)+H^{-1}_{3}\left(dx^{2}_{3}+dx^{2}_{4}\right)\right. \nonumber\\ \left. +H^{-1}_{2}\left(dx^{2}_{5}+dx^{2}_{6}\right)+d\tilde{s}^2\,\right]
\end{eqnarray}

\noindent
The four-dimensional base metric $d\tilde{s}^2$ is no longer Gibbons-Hawking but becomes:

\begin{equation}
\label{eq:GHAnsatz2}
d\tilde{s}^2 = 3^{-1}c^{-\frac{1}{3}}\left(H^{1} H^{2} H^{3}\right)^{-\frac{1}{3}} d\Psi^2 + 3c^{\frac{1}{3}}\left(H^{1} H^{2} H^{3}\right)^{\frac{1}{3}}\delta_{i j} dy^{i} dy^{j}\, , \qquad i = 1,2,3\,.
\end{equation}

This metric is not  Ricci flat and it does not even have constant curvature. This is to be expected, since the solution (\ref{eq:M2M2M2electricAnsatz}) solves the equations of motion of eleven-dimensional supergravity modified by appropriate higher curvature terms, which modify in turn the Gibbons-Hawking character of the base. Indeed, if one tries to compare this metric to a Gibbons-Hawking one, one finds that on one hand the gauge field $A_{i}(y)$, corresponding to D6 brane charge in ten dimensions, is zero, but that on the other hand the corresponding warp factor is not constant by rather has the form:

\begin{equation}
V(y) = 3c^{\frac{1}{3}}\left(H^{1} H^{2} H^{3}\right)^{\frac{1}{3}}\, , 
\label{AVnonGH}
\end{equation}

\noindent 
which has the same behavior at infinity and near the black hole as the warp factor of a Taub-NUT space with a nontrivial charge:

\begin{equation}
\lim_{r\to 0} V(y) \sim \frac{1}{r} \, ,
\end{equation}

\noindent
where $r=\sqrt{\left( y^{1}\right)^2 + \left( y^{2}\right)^2 + \left( y^{3}\right)^2}$. Furthermore, when the three warp factors become equal the function $V$ becomes harmonic throughout the space

\begin{equation}
\label{eq:deltaV}
\Delta_{y}\, V(y) = 0 \, ,
\end{equation}

\noindent despite the absence of a Gibbons-Hawking (D6) charge.

\noindent
It important to notice that in the M-theory uplift the term $ds^2_{\Psi\Psi}$ is constant, which is consistent with the ten-dimensional dilaton of the quantum black hole solution being constant. This is an nontrivial check that the eleven-dimensional brane configuration we propose gives the fundamental constituents of Type-IIA quantum black holes.


\section{The microscopic entropy}
\label{sec:microscopic}


Having obtained an eleven-dimensional metric that resembles that of three stacks of coincident M2 branes, we can easily compactify it along one of the torus directions to obtain a D2-D2-F1 metric, which upon a further T-duality along the F1 direction becomes a D3-D3-P metric, where the momentum P runs along the direction common to the two D3 branes. This duality chain transforms the quantum eleven-dimensional black hole into a type IIB D3-D3-P black hole, whose microscopic entropy can be reproduced straightforwardly by Strominger-Vafa-type arguments. Indeed, if the numbers of the two types of D3 branes are $N_1$ and $N_2$ and if $N_P$ units of momentum are running along the common directions of these branes, the most efficient way to carry this momentum when $N_1$ and $N_2$ are co-prime is to use the strings stretched between the two stacks of D3 branes. These ``bi-fundamental'' strings have a mass gap equal to $1\over N_1 N_2 R$, where $R$ is the radius of the common direction of the branes, and hence the entropy of the system comes from partitioning the $N_P$ units of momentum between modes that carry integer multiples of $1\over N_1 N_2$, or otherwise from the number of integer partitions of $N_1N_2 N_P$. By taking into account the fact that there are four bosonic species of bi-fundamentals as well as their fermionic partners, this gives the entropy $2 \pi \sqrt{N_1 N_2 N_P}$. 

The argument above reproduces the entropy of a D3-D3-P black hole in five dimensions, which is sourced by a stack of branes in $\mathbb{R}^4$. One can write this   $\mathbb{R}^4$ as a Gibbons-Hawking space with $V=1/r$ and a nontrivial fiber satisfying  $d A = \star dV$. If we focus instead on a stack of D3-D3-P branes in a Taub-NUT space with Kaluza-Klein monopole charge $N$, the corresponding warp factor is $V=1+\frac{N}{r}$ and we obtain a regular four-dimensional black hole with entropy $2 \pi \sqrt{N_1 N_2 N_P N}$ \cite{Johnson:1996ga,Maldacena:1996gb}.

We would like to use a similar argument to explain microscopically the entropy of our quantum black holes. However, at first glance there are two problems with this:
The first is that our base, (\ref{AVnonGH}), is no longer Gibbons-Hawking but has $A=0$ and $V= 3 c^{1\over 3}(H^1 H^2 H^3)^{1/3}$. Nevertheless, near the branes the warp factor behaves like that of a Gibbons-Hawking space. Hence, even if the branes do not sit on top of an $A_{N}$ singularity, they sit on some other conical singularity whose effect on the central charge of the D3-D3 CFT one can calculate. 

The second problem is that a generic eleven-dimensional uplift of a quantum black hole will not be a six-torus but a more complicated CY manifold. The key ingredient needed to relate the quantum black holes to the D3-D3-P system is the presence of two $U(1)$ isometries, one of which is used for reducing to a ten-dimensional Type IIA black hole, and the other for T-dualizing \footnote{If these two isometries are not present our microscopic description does not work, but neither does the microscopic description of ``normal'' M2-M2-M2 black holes.}. This can be easily done for any CY manifold that has a $T^2$ fiber. Nevertheless, in order for our construction of quantum black holes to yield regular solution, this elliptic fibration must be singular: CY manifolds with regular fibration have zero Euler characteristic and hence $c$ vanishes which makes the black hole horizon singular. This singularity can be cured by including non-perturbative $\alpha^{\prime}$ effects, again at tree-level in $g_{S}$ \cite{Bueno:2013psa}, but the resulting solutions involve the Lambert W function and are much harder to manage.

The way out is to focus on singular elliptic fibrations, which give CY manifolds  with nonzero Euler number. The places where the fibration degenerates become seven-branes upon dualization to the type IIB duality frame. The presence of these seven-branes does not affect the entropy counting, because this entropy comes from D3-D3 strings carrying momentum, which do not see the seven-branes. 

There are two ways to take into account the effect of the conical singularity on the entropy. The first is to compare this singularity with a Gibbons-Hawking solution, and determine its effective Gibbons-Hawking charge. The second is to focus on the near-horizon geometry of the black hole and to compute the corresponding Brown-Henneaux central charge \cite{Brown:1986nw}, which determines how the central charge of the D3-D3 CFT increases when the D3 branes are placed on top of the conical singularity. As we will explain below, the two calculations are equivalent, but since the first is more intuitive we will present it here.

Near the tip of a Gibbons-Hawking metric

\begin{equation}
\label{eq:GH3}
ds^2_{\mathrm{GH}} = V(y)^{-1}\left(d\Psi + A_{i}(y)dy^{i}\right)^2 + V(y)\left(dr^2 + d\Omega^2_{(2)}\right)\, , \qquad i = 1,2,3\, ,
\end{equation}

\noindent with 

\begin{equation}
\label{eq:GHconditions2}
V = 1 + \frac{N}{r}
\end{equation}

\noindent the metric becomes that of $\mathbb{R}^4/\mathbb{Z}_{N}$ 

\begin{equation}
\label{eq:GH-tip}
ds^2_{\mathrm{GH}}\sim d\rho^2 + \rho^2 d\tilde{\Omega}^2_{(3)}\, ,
\end{equation}

\noindent
with $\rho = 2\sqrt{r}$ and $d\tilde{\Omega}^2_{(3)}$ the standard metric on $S^3/\mathbb{Z}_{N}$. 
When the D3-D3 system is placed at the tip of this space its central charge increases by a factor of $N$. This is a well-known phenomenon for the D1-D5-P black hole in Taub-NUT \cite{Kutasov:1998zh}, and our system is just its T-dual. Now, given a conical metric of the type (\ref{eq:GH-tip}), there is a way to extract directly this factor:

\begin{equation}
\label{eq:N}
N = \frac{V_{S^3}}{V_{S^3/\mathbb{Z}_{N}}}\, ,
\end{equation}

\noindent
where $V_{S^3}$ is the volume of the three-sphere $S^{3}$ and $V_{S^3/\mathbb{Z}_{N}}$ is the volume of the $S^3/\mathbb{Z}_{N}$, at the same radius. In the Brown-Henneaux formalism this ratio of the volumes also gives the decrease of the effective three-dimensional Newton's constant, and hence the increase of the central charge of the corresponding CFT. 

For the quantum black hole metric we discussed in section \ref{sec:quantumM2}, the base metric

\begin{equation}
\label{eq:GHGfinal}
d\tilde{s}^2 = V(r)^{-1} d\Psi^2 + V(r)\left( dr^2 + r^2 d\Omega^2_{(2)}\right)\, ,\quad  V(y) = 3c^{\frac{1}{3}}\left(H^{1} H^{2} H^{3}\right)^{\frac{1}{3}}\, ,\quad i = 1,2,3
\end{equation}

\noindent
has a conical singularity in the near-tip region:

\begin{equation}
d\tilde{s}^2 \sim \frac{\sqrt{2}r}{3\left( c p^{1} p^{2} p^{3}\right)^{1/3}} d\Psi^2 + \frac{3\left( c p^{1} p^{2} p^{3}\right)^{1/3}}{\sqrt{2}r}\left( dr^2 + r^2 d\Omega^2_{(2)}\right)\, .
\end{equation}

\noindent Upon defining (with hindsight) 

\begin{equation}
\label{Neff}
N_E\equiv  \frac{\sqrt{3}\left( \sqrt{c}\, p^{1} p^{2} p^{3}\right)^{1/3}}{\sqrt{2}}\, ,
\end{equation}

\noindent 
and changing coordinates to ($\rho = 2\sqrt{N_E r}$), the metric near the conical singularity  becomes

\begin{equation}
\label{eq:GHGasymp}
 d\tilde{s}^2 \sim  d\rho^2 + \rho^2 \left(\frac{d\Psi^2}{4N^2_{E}}  +   \frac{d\Omega^2_{(2)}}{4} \right) \equiv d\rho^2 + \rho^2 d \Omega_{\rm transverse}\, .
\end{equation}

\noindent 
Hence, the singularity will increase the central charge of the D3-D3-P CFT by a factor given by the ratio of the volume of $S^3$ and the volume of the transverse space, which is nothing but $N_E$.

\begin{equation}
\label{eq:Neff}
 \frac{V_{S^{3}}}{V_{\Omega_{\rm transverse}}} = {2 \pi^2 \over {2 \pi \over 2 N_E} {4 \pi \over 2 }} = N_E\, .
\end{equation}


\noindent
Hence, the microscopic entropy of the quantum black holes will be given by

\begin{equation}
\label{entropy}
S = 2 \pi \sqrt{N_1 N_2 N_P N_E} \, ,
\end{equation}

\noindent
where $N_1$ and $N_2$ are the numbers of D3 branes and $N_P$ is the number of momentum quanta. Since 
the supergravity charges are proportional to these numbers, it is clear that this microscopic entropy count reproduces the correct charge growth of the quantum black hole

\begin{equation}
\label{entropymacro}
S = {3 \over 2}c^{\frac{1}{3}} \pi\left| {p}^1 {p}^2 {p}^3\right|^{2/3}\, ,
\end{equation}

\noindent which is already an important confirmation that our strategy is correct. Of course, the ideal would be to find exactly the coefficients that relate the supergravity charges to the quantized ones, and therefore establish that the macroscopic and the microscopic entropies are identical. However, since we  know neither the eleven-dimensional uplifts of the Maxwell fields of the four-dimensional quantum black hole, nor the volume of the 
Calabi-Yau three-fold and its submanifolds, we cannot determine these coefficients from first principles. However, what we can do is to use the symmetry of the STU model in order to argue that in the M2-M2-M2 duality frame the supergravity charges are related to the quantized brane numbers via the same proportionality constant, $\gamma$:

\begin{equation}
\label{eq:prewritting}
\tilde{p}^{1} = \gamma N_{1}\, , \qquad \tilde{p}^{2}  = \gamma N_{2} \, , \qquad \tilde{p}^{3}  = \gamma N_{P}\, ,
\end{equation}

\noindent
where we have defined $\tilde{p}^{i} \equiv \frac{p^{i}}{\sqrt{2}} $ to ease the presentation. We can now ask what is the value of $\gamma$ that makes the microscopic and the macroscopic entropies agree. Upon using \eqref{eq:prewritting}, equation \eqref{entropymacro} becomes:

\begin{equation}
\label{entropymacro2}
S = 2\pi \sqrt{\frac{3^{\frac{3}{2}}}{4} c^{\frac{1}{2}} \tilde{p}^{1}\tilde{p}^{2}\tilde{p}^{3} 3^{\frac{1}{2}}c^{\frac{1}{6}}\left(\tilde{p}^{1}\tilde{p}^{2}\tilde{p}^{3}\right)^{\frac{1}{3}}} = 2\pi \sqrt{\frac{3^{\frac{3}{2}}}{4} c^{\frac{1}{2}} \gamma^3 N_{1} N_{2} N_{P} N_{E}}
\end{equation}

\noindent
and therefore the desired value of $\gamma$ is:

\begin{equation}
\label{eq:gamma}
\gamma \equiv \frac{4^{\frac{1}{3}}}{3^{\frac{1}{2}}} c^{-\frac{1}{6}}\, .
\end{equation}

\noindent
There is a nontrivial check that this value satisfies: The number $N_{E}$, which gives the increase of the CFT central charge is expected to be a natural number, at least for some values of  $N_{1}$, $N_{2}$ and $N_{P}$. However, $N_{E}$ is defined in terms of $c$, and therefore contains both a factor of $\zeta(3)$ as well as square and cubic roots:

\begin{equation}
\label{eq:NE2}
N_{E} = \sqrt{3}\left( \sqrt{c}\, \tilde{p}^{1} \tilde{p}^{2} \tilde{p}^{3}\right)^{1/3}\, .
\end{equation} 

\noindent Hence one may naively infer that $N_{E}$ can never be a natural number. However, it turns out that when expressing $N_E$ in terms of quantized charges using \eqref{eq:gamma} both the $\zeta(3)$ and the $\sqrt{3}$ drop out:

\begin{equation}
\label{eq:NE3}
N_{E} =  \left( 4\, N_{1} N_{2} N_{P}\right)^{1/3}\, 
\end{equation} 

\noindent
and therefore $N_{E}$ can easily be an integer for a suitable choice of $N_1, N_2$ and $N_P$. 
The fact that the transcendental number in $c$ drops out of this relation is a nontrivial check of our proposed microscopic description. Indeed, in \cite{Behrndt:1997gs} it was argued that a black hole entropy expression that contains such a transcendental number can never be reproduced from microscopic calculations, and our microscopic proposal evades this by absorbing this transcendental number into the central charge increase of the underlying CFT. 

The puzzling aspect of equation (\ref{eq:NE3}) is that it makes the ``classical'' limit very hard to see. Indeed, as we explained in section (\ref{sec:IIAquantumblackholes}) if one turns off the quantum corrections ($c\rightarrow 0$) while keeping the {\em supergravity} charges of the black hole constant, the horizon becomes singular. This is reflected in our construction by the fact that 
our black hole has three charges and, when $c\rightarrow 0$, the warping of the base space becomes trivial and the corresponding entropy becomes that of the three-charge system in four dimensions, which does not give rise to a macroscopic horizon.
 
However, one can also ask what happens if one takes the $c\rightarrow 0$ limit while keeping the quantized black hole charges, $N_1, N_2$ and $N_P$ constant. This does not affect $N_E$, so it looks like in this limit  the black hole entropy remains macroscopic. This is consistent with the fact that in this limit the four-dimensional supergravity charges blow up (\ref{eq:gamma}), but the factors of $c$ cancel from the expression of the near-horizon limit of the metric (\ref{eq:generalbhmetricsusySTU}), which remains a regular $AdS_2 \times S^2$. On the other hand, the expressions of the four-dimensional moduli diverge, and therefore it appears that in this limit the  dictionary between ten- and four-dimensional solutions breaks down.
 It would clearly be interesting to try to derive equation (\ref{eq:gamma}) from first principles to see precisely how this breakdown occurs. 


\section{Conclusions}
\label{sec:conclusions}


We have constructed an eleven-dimensional metric (\ref{eq:M2M2M2electricAnsatz}) that upon dimensional reduction gives the Type-IIA quantum black hole of the STU quantum-corrected model. Because the four-dimensional metric has constant dilaton and CY volume, this eleven-dimensional uplift can be interpretation as arising from three stacks of orthogonal M2 branes that sit at the apex of a cone in a four-dimensional transverse space. Because of the presence of correction terms in the Lagrangian, this space is not Gibbons-Hawking, although it has exactly the same kind of warping as a Gibbons-Hawking space. The strength of the conical singularity is proportional to the cubic root of the product of the three M2 charges

When the dix-dimensional internal space of the compactification has a $T^2$ fiber we can dualize this solution to an asymptotically four-dimensional three-charge D3-D3-P solution in type IIB string theory. In flat space the microscopic entropy of this system is not enough to give rise to a regular horizon, but we have shown that the conical singularity enhances the central charge of this system, and the resulting microscopic entropy reproduces the entropy of the of the quantum black hole: $S = \frac{3c^{\frac{1}{3}}}{2} \pi\left| p^1 p^2 p^3\right|^{2/3}$ up to an overall coefficient which we could not determine. We have however been able to show that if this coefficient is such that the entropies match, a certain dependence of the entropy on transcendental numbers drops out, which we believe is a nontrivial check of our proposal. 

Since we are working in a supergravity theory in the presence of quantum corrections to the geometry of the scalar manifold, our proposed microscopic description is not at the same level of rigor as the usual three- and four-charge black hole entropy counting. However, the fact that we have found a brane interpretation that reproduces the highly-unusual charge dependence of the entropy of quantum black holes makes us confident that we have identified the correct microscopic framework for understanding the entropy of all Type-IIA quantum black holes, which remains as an important open problem in String Theory.


\acknowledgments

We would like to thank T. Ort\'in for useful discussions and comments. The work of IB was supported in part by the ERC Starting Independent Researcher Grant 240210, String-QCD-BH, by the John Templeton Foundation Grant 48222: ``String Theory and the Anthropic Universe'' and by a grant from the Foundational Questions Institute (FQXi) Fund, a donor advised fund of the Silicon Valley Community Foundation on the basis of proposal FQXi-RFP3-1321 to the Foundational Questions Institute.The work of CS was supported by the ERC Starting Independent Researcher Grant 259133, ObservableString.

\appendix


\section{$\mathcal{N}=2$ four-dimensional supergravity and the H-FGK formalism}
\label{sec:N2SugraHFGK}


Type-IIA quantum black holes are black hole solutions of Type-IIA String Theory compactified down to four dimensions on a Calabi-Yau three-fold, which is described, up to two derivatives, by a $\mathcal{N}=2$ four-dimensional ungauged supergravity. Therefore, it is convenient to review the basic formulation of the theory and its vector multiplet sector, since the hypermultiplets and the fermions can be always truncated for black hole solutions. The bosonic sector of any $\mathcal{N}=2$ four-dimensional supergravity  coupled to vector multiplets can be written as follows \cite{deWit:1984px,Andrianopoli:1996cm}

\begin{equation}
\label{eq:action}
S= \int d^{4}x \sqrt{|g|}\,
\left\{
R
+\mathcal{G}_{i\bar{j}}(z,\bar{z})\partial_{\mu}z^{i}\partial^{\mu}\bar{z}^{\bar{j}} +2I_{\Lambda\Sigma}(z,\bar{z})F^{\Lambda}{}_{\mu\nu}F^{\Sigma\, \mu\nu}-2R_{\Lambda\Sigma}(z,\bar{z})F^{\Lambda}{}_{\mu\nu} \star F^{\Sigma\, \mu\nu}
\right\}\,
\end{equation}

\noindent
The $z^i$ ($i=1,...n_v$) denote the $n_v$ complex scalar fields of the vector multiplets, which parametrize an $n_v$-dimensional Special K\"ahler manifold with K\"ahler metric $\mathcal{G}_{i\bar{j}}(z,\bar{z})$. $F^{\Lambda} = dA^{\Lambda}$ denote the field strengths of the $\Lambda=0,...,n_v$ one-form connections $A^{i}$ that belong to the vector multiplets, plus the graviphoton $A^0$. The real matrices $I _{\Lambda\Sigma}\equiv \mathrm{Im}\,\mathcal{N}_{\Lambda\Sigma}(z,\bar{z})$, $R_{\Lambda\Sigma} \equiv \mathrm{Re}\, \mathcal{N}_{\Lambda\Sigma}(z,\bar{z})$ denote respectively the imaginary, negative definite, and real parts of the symplectic complex \emph{period} matrix $\mathcal{N}$. Hence, the period matrix determines the couplings of the one-form connections $A^{\Lambda}$  to the scalars $z^{i}$ of the vector multiplets. The equations of motion following from the action (\ref{eq:action}) are given by

\begin{eqnarray}
G_{\mu\nu}
+2\mathcal{G}_{i\bar{j}}[\partial_{\mu}z^{i} \partial_{\nu}\bar{z}^{\bar{j}}
-{\textstyle\frac{1}{2}}g_{\mu\nu}
\partial_{\rho}z^{i}\partial^{\rho}\bar{z}^{\bar{j}}]
+8\mathrm{Im}\,\mathcal{N}_{\Lambda\Sigma}
F^{\Lambda\, +}{}_{\mu}{}^{\rho}F^{\Sigma\, -}{}_{\nu\rho} = 0\, ,
\label{eq:Emn}\\
\nonumber \\
\nabla_{\mu}(\mathcal{G}_{i\bar{j}}
\partial^{\mu} \bar{z}^{\bar{j}})
-\frac{1}{2}\partial_{i}\mathcal{G}_{j\bar{k}}\partial_{\rho}z^{j}
\partial^{\rho}\bar{z}^{\bar{k}} 
+\frac{1}{2}\partial_{i}[
G_{\Lambda}{}^{\mu\nu}\ast F^{\Lambda}{}_{\mu\nu}] = 0\, ,
\label{eq:Ei}\\
\nonumber \\
\nabla_{\nu}\ast G_{\Lambda}{}^{\nu\mu} = 0\, ,
\label{eq:ERm}
\end{eqnarray}

\noindent
where we have defined

\begin{equation}
\label{eq:defGmag}
G_{\Lambda} \equiv   
-\frac{1}{4\sqrt{-g}}\frac{\delta S}{\delta \ast F^{\Lambda}}
= \mathrm{Re}\,\mathcal{N}_{\Lambda\Sigma}F^{\Sigma} +\mathrm{Im}\,\mathcal{N}_{\Lambda\Sigma}\ast F^{\Sigma}\, .
\end{equation}

\noindent
The Maxwell equations for the field strengths $F^{\Lambda}$ together with the corresponding Bianchi identities can be written in terms of differential forms as follows

\begin{equation}
\label{eq:Maxwell}
dG_{\Lambda} = 0\, , \quad dF^{\Lambda} = 0\, .
\end{equation}

\noindent
Notice that, since $G_{\Lambda}$ is a closed two-form, it can be written locally as the exterior derivative of a one-form  $A_{\Lambda}$

\begin{equation}
G_{\Lambda} = dA_{\Lambda}\, .
\end{equation}

\noindent
where $A_{\Lambda}$ is the so-called magnetic dual of $A^{\Lambda}$ and both sets of connection one-forms can be arranged into a symplectic vector $A^{M} = \left( A^{\Lambda}, A_{\Lambda}\right)^{T}$. Supersymmetry constrains the couplings of all the fields of the theory in a very precise way which, for the vector-multiplet sector, is elegantly encoded in the language of Special K\"ahler Geometry \cite{deWit:1983rz,deWit:1984px}. In fact, the bosonic Lagrangian of $\mathcal{N}=2$ four-dimensional supergravity coupled to vector multiplets is determined by choosing a holomorphic section $\Omega\in \Gamma\left(\mathcal{SV}\right)$ or, equivalently (when it happens to exist), a homogeneous function $\mathcal{F}\left(\mathcal{X}\right)$ of degree two, the $\mathcal{N}=2$ \textit{prepotential}, from which $\mathcal{G}_{i\bar{j}}$ and $\mathcal{N}_{\Lambda\Sigma}$ can be easily obtained as

\begin{eqnarray}
\mathcal{G}_{i\bar{j}}&=&-\partial_i\partial_{\bar{j}} \mathrm{log}\left\{i \left[\bar{\mathcal{X}}^{\Lambda}\partial_{\Lambda}\mathcal{F}-\mathcal{X}^{\Lambda}\partial_{_{\Lambda}}\bar{\mathcal{F}} \right] \right\}\, ,\\
\mathcal{N}_{\Lambda\Sigma}&=&\partial_{\Lambda\Sigma}\bar{\mathcal{F}}+2i\frac{\mathrm{Im} (\partial_{\Lambda\Lambda^{\prime}}\mathcal{F})\mathcal{X}^{{\Lambda}^{\prime}}\mathrm{Im} (\partial_{\Sigma\Sigma^{\prime}}\mathcal{F})\mathcal{X}^{{\Sigma}^{\prime}}}{\mathcal{X}^{\Omega}\mathrm{Im}(\partial_{\Omega\Omega^{\prime}}\mathcal{F})\mathcal{X}^{\Omega^{\prime}}}\, . \,\,\, \,\,\, \,\,\, \,\,
\end{eqnarray}

\noindent
Here $\mathcal{X}^{\Lambda}$ denote the homogeneous coordinates on the scalar manifold, related to the scalar fields $z^i$ via

\begin{equation}
z^{i} \equiv \frac{\mathcal{X}^{i}}{\mathcal{X}^0}\, ,
\end{equation}

\noindent
Therefore, choosing a second-degree homogeneous function $\mathcal{F}\left(\mathcal{X}\right)$ automatically determines an $\mathcal{N}=2$, four-dimensional, ungauged supergravity theory coupled to vector multiplets, which has the appropriate matter content for constructing black hole solutions. The most general static and spherically symmetric metric that solves the equations of motion (\ref{eq:action}) is given by \cite{Shahbazi:2013ksa,Ferrara:1997tw,Meessen:2011aa}

\begin{equation}
\label{eq:generalbhmetric}
\begin{array}{rcl}
ds^2_{4}
& = &
-e^{2U(\tau)} dt^{2} + e^{-2U(\tau)} \gamma_{\underline{m} \underline{n}}
dx^{\underline{m}} dx^{\underline{n}}\, ,  \\
& & \\
\gamma_{\underline{m}\underline{n}}
dx^{\underline{m}} dx^{\underline{n}}
& = & 
\frac{r_{0}^{2}}{\sinh^{2} r_{0}\tau}\left[ \frac{r_{0}^{2} }{ \sinh^{2}r_{0}\tau} d\tau^2
+ d\theta^2 +\sin^2\theta d\phi^2\right]\, , 
\end{array}
\end{equation}

\noindent
where $\tau$ is the radial coordinate. When (\ref{eq:generalbhmetric}) describes a physical black hole solution, $r_0$ is the \textit{non-extremality parameter}, the exterior of the event horizon corresponds to  $\tau\in(-\infty,0)$; the event horizon is at $\tau = -\infty$ and spatial infinity corresponds to $\tau\rightarrow 0^-$. The inner part of the Cauchy horizon corresponds to $\tau\in(\tau_s,\infty)$, with the inner horizon at $\tau\rightarrow \infty$ and the singularity at $\tau=\tau_s$ for a certain positive and finite real number $\tau_s$ \cite{Galli:2011fq}. Since the metric is spherically symmetric, we will assume that all the fields of the theory depend exclusively on the radial coordinate $\tau$. We define the black hole charges as

\begin{equation}
\label{eq:chargesFG}
p^{\Lambda} = \int_{S^{2}_{\infty}} i^{\ast}F^{\Lambda}\, , \quad q_{\Lambda} = \int_{S^{2}_{\infty}} i^{\ast}G_{\Lambda}\, ,
\end{equation}

\noindent
where $S^{2}_{\infty}$ denotes an space-like two-sphere at spatial infinity $\tau\to 0$, $p^{\Lambda}$ correspond to the magentic charges and $q_{\Lambda}$ correspond to the electric charges of the black hole, which can be together arranged into a symplectic vector $\mathcal{Q}^M\equiv\left(p^{\Lambda}, q_{\Lambda} \right)^T$. In the background given by (\ref{eq:generalbhmetric}) Maxwell's equations can be integrated explicitly, in such a way that the complete electric connection one-form $A^{\Lambda}$ is given in terms of the time component $A^{\Lambda}_{t}$ of the electric connection one-form and the time component of the magnetic connection one-form $A_{\Lambda\, t}$. Indeed, let $\Sigma^{M} \equiv ( A^{\Lambda}_{t},A_{\Lambda\, t})^{T}$ be a symplectic vector made from the time components of the electric $A^{\Lambda}$ and magnetic $A_{\Lambda}$ connection one-forms. Then, it can be shown that

\begin{equation}
\label{eq:vectorialesfields}
\Sigma^{M} = \frac{1}{2}\int e^{2U} \mathcal{M}^{MN}\mathcal{Q}_{N}d\tau\, ,
\end{equation}

\noindent
where $\mathcal{M}_{MN}$ is a symplectic and symmetric matrix constructed from the couplings of the scalars and the vector fields as

\begin{equation}
\label{eq:M}
\mathcal{M}_{MN}\left(\mathcal{N}\right) 
\equiv
\left(
\begin{array}{cc}
\left( I+R I^{-1}R\right)_{\Lambda\Sigma}  & -\left( R I^{-1}\right)_{\Lambda}{}^{\Sigma} \\
& \\
-\left(I^{-1} R\right)^{\Lambda}{}_{\Sigma} & \left( I^{-1} \right)^{\Lambda\Sigma} \\
\end{array}
\right)\, .
\end{equation}

\noindent
We choose to express all Maxwell field strengths in terms of the time components of the electric and the magnetic connection one-forms. For the electric field strengths this gives:

\begin{equation}
\label{eq:FLambda}
F^{\Lambda}_{t\tau} = -\partial_{\tau} A^{\Lambda}_{t}\, , \quad F^{\Lambda}_{\theta\phi} = \sin\theta\, e^{-2U}\left(\left(I^{-1}\right)^{\Lambda\Sigma}\frac{dA_{\Sigma\, t}}{d\tau} - \left(I^{-1}R\right)^{\Lambda}{}_{\Sigma} \frac{dA^{\Sigma}_{t}}{d\tau}\right) \, ,
\end{equation}

\noindent
and the expression for the magnetic field strengths $G_{\Lambda}$ can be similarly obtained from equation (\ref{eq:FLambda}) using equation (\ref{eq:defGmag}). Since the connection one-forms can be explicitly integrated, they can be eliminated from the action. The four-dimensional $\mathcal{N}=2$ ungauged supergravity action coupled to vector multiplets can then be shown to be completely equivalent, assuming the space-time background given (\ref{eq:generalbhmetric}) and radial dependence for all the fields, to the one-dimensional effective FGK action \cite{Ferrara:1997tw} for the $2 n_{v}$ complex fields $z^i(\tau)$ and the real field $U(\tau)$

\begin{equation}
\label{eq:FGK}
S_{\mathrm{FGK}}\left[U,z\right]=\int d\tau \left\{ \dot{U}^2+\mathcal{G}_{i\bar{j}}\dot{z}^i\dot{\bar{z}}^{\bar{j}}-e^{2U}V_{\rm bh}(z,\bar{z},\mathcal{Q}) \right\}\, ,
\end{equation}

\noindent
together with the \textit{Hamiltonian constraint}, 

\begin{equation}
\label{eq:haml}
\dot{U}^2+\mathcal{G}_{i\bar{j}}\dot{z}^i\dot{\bar{z}}^{\bar{j}}+e^{2U}V_{\rm bh}(z,\bar{z},\mathcal{Q}) = r^{2}_{0}\, .
\end{equation}

\noindent
Here $V_{\rm bh}$ is the so-called \textit{black hole potential}, which is given by \cite{Ferrara:1997tw}

\begin{eqnarray}
\label{VBH}
V_{\rm bh}(z,\bar{z},\mathcal{Q})&\equiv& \frac{1}{2}\mathcal{M}_{MN}(\mathcal{N})\mathcal{Q}^M\mathcal{Q}^N\, .
\end{eqnarray}

\noindent
We are now ready to introduce the H-FGK formalism. The H-FGK formalism \cite{Galli:2011fq,Meessen:2011aa,Mohaupt:2011aa,Galli:2012ji,Bueno:2013pja} consists of a particular change of variables from the $(2n_v+1)$-real $\left(U,z^{i}\right)$ to a new set of $(2n_v+2)$-real variables $H^M(\tau)$ which transform as a symplectic, linear, representation the U-duality group of the theory, and become harmonic functions in Euclidean $\mathbb{R}^3$ in the supersymmetric solution.  The equations of motion in the new variables $H^{M}(\tau)$ can be written as

\begin{eqnarray}
\label{eq:Eqsmotion}
\frac{1}{2}\partial_{PMN}\log \mathsf{W}\,
\left[\dot{H}^{M}\dot{H}^{N} -\frac{1}{2}\mathcal{Q}^{M}\mathcal{Q}^{N}
\right]
+ \partial_{PM}\log \mathsf{W}\, \ddot{H}^{M}
-\frac{d}{d\tau}\left(\frac{\partial \Lambda}{\partial \dot{H}^{P}}\right)
+\frac{\partial \Lambda}{\partial H^{P}}=0\, 
\end{eqnarray}

\noindent
together with the \emph{Hamiltonian constraint}

\begin{eqnarray}
\label{eq:Hamconstraint}
-\frac{1}{2}\partial_{MN}\log\mathsf{W}
\left(\dot{H}^{M}\dot{H}^{N}-\frac{1}{2}\mathcal{Q}^{M}\mathcal{Q}^{N} \right)
+\left(\frac{\dot{H}^{M}H_{M}}{ \mathsf{W}}\right)^{2}
- \left(\frac{\mathcal{Q}^{M}H_{M}}{ \mathsf{W}}\right)^{2}
-r_{0}^{2}=0\, 
\end{eqnarray}

\noindent
where
\begin{equation}
\Lambda \equiv \left(\frac{\dot{H}^{M}H_{M}}{ \mathsf{W}}\right)^{2}
+\left(\frac{\mathcal{Q}^{M}H_{M}}{ \mathsf{W}}\right)^{2}\, ,
\end{equation}

\noindent
and
\begin{equation}
\label{eq:W(H)}
e^{-2U} = \mathsf{W}(H) \equiv \tilde{H}_{M}(H)H^{M}\, , \qquad \tilde{H}^M+iH^M =\frac{\mathcal{V}^M}{X}\, ,
\end{equation}

\noindent
with $\mathcal{V}^M$ being the covariantly holomorphic symplectic section that determines the vector-multiplet sector of $\mathcal{N}=2$ supergravity, and $X$ a complex variable with the same K\"ahler weight as $\mathcal{V}^M$, making the quotient $\mathcal{V}^M/X$ K\"ahler invariant. The symplectic vector $\tilde{H}^M(\mathcal{I})\equiv \tilde{H}^M(H)$ stands for the real part of $\mathcal{V}^M$ written as a function of the imaginary part, $H^M$; this can always be done by solving the  \emph{stabilization equations}. The function $\mathsf{W}(H)$ is usually known in the literature as the \emph{Hesse potential}.

The effective theory is now expressed in terms of $2\left(n_{v}+1\right)$ variables $H^M$. The solution depends on $2\left(n_{v}+1\right)+1$ parameters, namely the $2\left(n_{v}+1\right)$ charges $\mathcal{Q}^M=\left(p^{\Lambda},\, \, q_{\Lambda}\right)^T$ and the non-extremality parameter $r_{0}$, from which it is always possible to reconstruct the complete solution in terms of the four-dimensional fields of the theory. The H-FGK formalism introduces an extra  real degree of freedom. Hence the H-FGK action enjoys an extra gauge symmetry which, by gauge fixing, allows to get rid of the extra degree of freedom \cite{Galli:2012ji}.


\section{A \emph{quantum} class of black holes}
\label{sec:quantumblackholes}


In this appendix we present the solution of the equations (\ref{eq:Eqsmotion}) and (\ref{eq:Hamconstraint}) that correspond to the quantum black holes of Type-IIA String Theory. Type-IIA quantum black holes are based on the following truncation of the $H$-variables and the charges

\begin{equation}
\label{eq:truncation}
H^{0} = H_{0} = H_{i} = 0\, , \qquad p^{0}=q_{0}=q_{i} = 0\, .
\end{equation}

\noindent
Using now equation (\ref{eq:truncation}) together with equations (\ref{eq:W(H)}) and (\ref{eq:prepotential}) we find

\begin{equation}
\label{eq:hessian}
e^{-2U} = \mathsf{W}(H)=\frac{\left( 3! c\right)^{1/3}}{2}\left|\kappa^{0}_{ijk} H^i H^j H^k\right|^{2/3}\, ,
\end{equation}

\noindent
where $c=\frac{\chi\zeta(3)}{ (2\pi)^3}$. From equation \eqref{eq:hessian} it follows that, in order to have a non-singular metric, $c$ must be positive, that is, $h^{1,1}>h^{2,1}$ is a necessary condition in order to obtain an admissible solution. There are plenty of Calabi-Yau manifolds that satisfy this condition, so we will not worry any more about it. The scalar fields, purely imaginary, are

\begin{equation}
\label{eq:scalars}
z^i= i \left(3!\right)^{\frac{1}{3}} c^{\frac{1}{3}} \frac{H^i}{\left(\kappa^{0}_{ijk} H^i H^j H^k\right)^{1/3}} \, ,
\end{equation}

\noindent
It is easy to see that the solution is not consistent in the classical limit $c\to 0$, and also that no classical limit can be assigned to it, since when $c=0$ the model is already singular before solving the equations of motion. Hence, we conclude that the corresponding solutions are \emph{genuinely} quantum solutions, i.e., they only exist when the perturbative quantum corrections are incorporated into the action, and thus they are called Type-IIA {\em quantum} black holes.

Of course, we still have to solve the $H^{i}\, , i=1,\cdots , n_{v},\,$ as functions of the radial coordinate $\tau$. Since in this letter we are interested only in supersymmetric solutions, we automatically know that \cite{Gauntlett:2002nw,Meessen:2006tu,Huebscher:2006mr}

\begin{equation}
\label{eq:universalsusy}
H^{i} = a^{i}-\frac{p^{i}}{\sqrt{2}}\tau\, , \qquad r_0=0\, ,
\end{equation}

\noindent
that is, the $H^{i}\, , i=1,\cdots , n_{v},\,$ are given by harmonic functions on $\mathbb{R}^{3}$. For supersymmetric solutions we have to take $r_{0}\to 0$ and therefore the general metric (\ref{eq:generalbhmetric}) simplifies to

\begin{equation}
\label{eq:generalbhmetricsusy}
\begin{array}{rcl}
ds^2_{4} = -\,
e^{2U(\tau)} dt^2 + e^{-2U(\tau)} \delta_{\underline{m} \underline{n}}
dx^{\underline{m}}dx^{\underline{n}}\, , 
\end{array}
\end{equation}

\noindent
where $\delta$ is the Euclidean metric on $\mathbb{R}^3$.  The near horizon $\tau\rightarrow -\infty$ limit of the metric (\ref{eq:generalbhmetricsusy}) is given by

\begin{equation}
\label{eq:nearhorizonsusy}
\begin{array}{rcl}
\lim_{\tau\rightarrow -\infty}ds^2_{4} = -
\frac{4}{\left( 3! c\right)^{1/3}}\left|\kappa^{0}_{ijk} p^i p^j p^k\right|^{-2/3} \tau^{-2} dt^2 + \frac{\left( 3! c\right)^{1/3}}{4}\left|\kappa^{0}_{ijk} p^i p^j p^k\right|^{2/3} \tau^2 \delta_{\underline{m} \underline{n}}
dx^{\underline{m}} dx^{\underline{n}}\, .
\end{array}
\end{equation}

\noindent
The entropy of the Type-IIA quantum black holes is given by

\begin{equation}
\label{eq:entropy}
S = \frac{\left( 3! c\right)^{1/3}}{4}\pi\left|\kappa^{0}_{ijk} p^i p^j p^k\right|^{2/3}\, .
\end{equation}

\noindent
As we explained in section \ref{sec:N2SugraHFGK}, the connection one-forms $A^{\Lambda}$ can be explicitly obtained using equation (\ref{eq:vectorialesfields}). For Type-IIA quantum black holes we obtain

\begin{equation}
R_{00}= 0\, , \qquad  I_{0i} = I_{i0} = 0\, , \qquad R_{ij} = 0\, ,
\end{equation}

\noindent
which in turn implies that the following components of $\mathcal{M}_{MN}$ are zero

\begin{equation}
\label{eq:Mcomponents}
\mathcal{M}_{0i} = \mathcal{M}_{i0}  = 0\, ,\qquad \mathcal{M}_{0}{}^{0} = \mathcal{M}^{0}{}_{0} = \mathcal{M}_{i}{}^{j} = \mathcal{M}^{i}{}_{j} = 0\, , \qquad \mathcal{M}^{0i} = \mathcal{M}^{i0} = 0\, .
\end{equation}

\noindent
From equations (\ref{eq:vectorialesfields}) and (\ref{eq:truncation}) we obtain

\begin{equation}
\label{eq:vectorfieldsII}
\Psi_{M} = \frac{1}{2}\int e^{2U} \mathcal{M}_{Mi} p^{i} d\tau\, ,
\end{equation}

\noindent
and since 

\begin{equation}
\Psi_{M} = \left( A_{\Lambda\, t}, - A^{\Lambda}_{\, t}\right)^{T}\, ,
\end{equation}

\noindent
we conclude using (\ref{eq:Mcomponents}) that the only non-zero components of $\Psi^{M}$ are

\begin{equation}
A_{i\, t} = \frac{1}{2}\int e^{2U} \mathcal{M}_{ij}\, p^{j} d\tau\, , \qquad A^{0}_{\, t} = - \frac{1}{2}\int e^{2U} \mathcal{M}^{0}{}_{j}\, p^{j} d\tau\, .
\end{equation}

\noindent
This implies that the connection one-forms $A^{i}$ have only magnetic components, which give rise to the magnetic charges $p^{i}$ of the black hole solution (\ref{eq:chargesFG}). Notice however that the time component of graviphoton $A^{0}$ is non-zero, although the corresponding charges are zero: the magnetic one  because $i^{\ast}F$ is identically zero and the electric one thanks to a precise cancellation in the corresponding formula for $q_{0}$:

\begin{equation}
q_{0} = \int_{S^{2}_{\infty}} i^{\ast}G_{0}\, .
\end{equation}


\renewcommand{\leftmark}{\MakeUppercase{Bibliography}}
\phantomsection
\bibliographystyle{JHEP}
\bibliography{References}
\label{biblio}

\end{document}